# On full text download and citation distributions in scientific-scholarly journals

Henk F. Moed * and Gali Halevi **

* Corresponding author. Informetric Research Group, Elsevier, Radarweg 29, 1043 NX Amsterdam (The Netherlands). Email: h.moed@elsevier.com.

** Informetric Research Group, Elsevier, 360 Park Av. South, New York, NY 10011 (USA). Email: g.halevi@elsevier.com.

## Abstract

A statistical analysis of full text downloads of articles in Elsevier's ScienceDirect covering all disciplines reveals large differences in download frequencies, their skewness, and their correlation with Scopus-based citation counts, between disciplines, journals, and document types. Download counts tend to be two orders of magnitude higher and less skewedly distributed than citations. A mathematical model based on the sum of two exponentials does not adequately capture monthly download counts. The degree of correlation at the article level within a journal is similar to that at the journal level in the discipline covered by that journal, suggesting that the differences between journals are to a large extent discipline-specific. Despite the fact that in all study journals download and citation counts per article positively correlate, little overlap may exist between the set of articles appearing in the top of the citation distribution and that with the most frequently downloaded ones. Usage and citation leaks, bulk downloading, differences between reader and author populations in a subject field, the type of document or its content, differences in obsolescence patterns between downloads and citations, different functions of reading and citing in the research process, all provide possible explanations of differences between download and citation distributions.

# 1. Introduction

In the past decade, many scientific literature publishers have implemented usage monitoring systems based on data including clickstreams, downloads and views of scholarly publications recorded on an article level, that allow them to capture the number of times articles are downloaded in their PDF or HTML formats. This type of data is not only used by publishers as a way to monitor the usage of their journals but also by libraries who wish to monitor and manage the usage of their collections (Duy & Vaughan, 2006).

The growing need for this type of monitoring resulted in the launch of COUNTER (Counting Online Usage of Networked Electronic Resources), an international initiative which aimed to set standards and facilitate the recording and reporting of online usage statistics in a consistent, credible and compatible way. Nowadays, COUNTER is an industry standard, used by most publishers and libraries and allows for downloads data to be analyzed and compared more easily by subscribers and publishers alike. This development could be one of the reasons that research in this area has seen such significant growth.

During the past decade, the relationship between citations and full text article downloads has gained significant attention. In their review article published in 2010, Michael Kurtz and Johan Bollen describe "Usage Bibliometrics "as the statistical analysis of how researchers access their technical literature, based on the records that electronic libraries keep of every user transaction (Kurtz & Bollen, 2010). They underline that many "classical", citation-based measures have direct analogs with usage, and that an important approach to validation of usage statistics is to demonstrate the similarities and differences between citation and usage statistics. An important class of usage statistics is based on the number of times articles from publication archives are downloaded in full text format, denoted as "downloads" below. Kurtz and Bollen claim that "….the relation between usage and citation has not been convincingly established" (p. 23) and that "….direct comparisons over the same set of input documents are rare" (p. 23).

Kurtz et al. (2005a; 2005b) published two pioneering papers analyzing usage mainly of the NASA Astrophysics Data System (ADS), and comparing the number of electronic accesses – which they term "reads" – of individual articles in astronomy and astrophysics journals with citation counts. They described the obsolescence patterns of download counts as the sum of three exponentials, representing three distinct usage modes: "historical", "interesting" and "current". A fourth mode, denoted as "new", relates to the usage of recently published journal issues and was left out of the model as its effect could not detected in their data.

Kurtz et al. (2005b) found that in their astronomy data the citation obsolescence function follows the usage function very closely. It has two components representing the interesting and current mode, respectively, of which the parameters are equal to those of the usage function. The functional relationship between citations and downloads is assumed to be essentially linear. They add a component of the form *[1-exp (-kt)]* in which *t* is the time variable and *k* expresses the delay of citations compared to downloads, which they ascribe to inefficiencies in the publication process. The linearity constant expresses the number of downloads ("reads" in the terminology of Kurtz et al.) per citation. They observed that its value depends upon the citation data base used, and on the overall increase in usage, but concluded that in their data it hardly changes with the age of the used articles; in this sense they assume it is a genuine constant.

After the publication of these articles, a series of articles explored statistical aspects of usage, including Perneger, 2004; Moed, 2005; Davis & Fromerth, 2007; O'Leary, D.2008; Schloegl & Gorraiz, 2011; Xue-li, Hong-ling & Mei-ying, 2011; Nieder, Dalhaug & Aandahl, 2013; Lippi & Favaloro, 2013; and Gorraiz, Gumpenberger & Schloegl, 2014. Several papers studied this relationship in order to develop predictive models of citations based on early usage figures (Broody, Harnad & Carr, 2006; Jahandideh, Abdolmaleki, & Asadabadi, 2007; Sharma, 2007; Zavos, 2008). Citations, publications and usage are combined in order to develop models that can capture the weight of each one and provide better understanding of the relationship between them and how those can be applied to an institution and individual's assessments (Bollen & Van De Sompel, 2008). Usage-based indicators are, jointly with measures of the number of mentions an article receives in social platforms (Barjak, et.al, 2007; Adie & Roe, 2013; Taylor, 2013) are sometimes labeled as "altmetrics".

The first author of the current paper published in 2005 an analysis of the statistical relationship between citations and full text article downloads for articles in one particular journal: Tetrahedron Letters, published by Elsevier (Moed, 2005). The paper examined the patterns of publications obsolescence using synchronous and diachronous (or asynchronous) approaches which were applied to the analysis of downloads and their recorded citations in the Science Citations Index (SCI). The analysis using a synchronous approach showed that journal download counts per month can be described by a model consisting of the sum of two negative exponential functions, representing an ephemeral and a residual factor, whereas the decline phase of citations conforms to a simple exponential function with a decay constant statistically similar to that of the downloads residual factor. A diachronous approach showed that, as a cohort of documents grows older, its download distribution becomes more and more skewed, and more statistically similar to its citation distribution. The article also presented a method aimed to estimate the effect of citations upon downloads using obsolescence patterns

A main objective of the current paper is to expand the analyses presented in the 2005 article in the following ways:

- Analyze a much larger set of journals covering all domains of science and scholarship, and highlight differences in downloading behavior between these domains.
- Give information on the order of magnitude of download and citation counts and the ratio of downloads and citations at the level of journals and document types.
- Provide insight into downloading practices of users, analyzing user sessions, institutions and countries.
- Analyze download obsolescence functions and time delays for a series of journals, based on monthly rather than annual counts.
- Compare the skewness of the download and citation article distributions.
- Examine the statistical correlation between downloads and citations both at the level of journals within a discipline and at that of individual articles within a journal.

The base assumption underlying this paper is that a sound statistical analysis of the relationship between downloads and citations, and a thorough reflection upon its outcomes, contributes to a better understanding of what both download counts and citation counts measure, or more generally, allow more insight into information retrieval, reading, and referencing practices in scientific-scholarly research. It is the very combination of the two types of data that expands, so to speak, the horizon, and provides a perspective in which each of the two types can be positioned more adequately. In the quantitative study of research activity and performance, downloads and citations provide complementary data sources.

In this article the term "usage" is reserved for the use made of electronic publication archives in the broadest sense, and recorded in the archive's electronic log files. It includes activities such as downloading in pdf, viewing in html format, browsing through abstracts, and also saving, sharing or annotating documents in reference managers. The article focuses on downloading the full text, either in PDF or in HTML format, of a document indexed in ScienceDirect, Elsevier's online full text article database. Throughout this paper the term "downloads" means: "full text article download (in PDF or HTML format) from ScienceDirect. The term "use" is a neutral term, indicating both usage as defined above, and citation counts in journals indexed in Scopus, and is mainly used in figures presenting both download and citation data on the same axis.

*The structure of this article* is as follows. Section 2 presents an overview of the main factors that one should keep in mind when interpreting full text article download counts and their relationship to citations in peer reviewed journal articles. The overview is partly based upon a literature review, especially the thorough review by Kurtz & Bollen (2010), but adds factors or examples based on empirical findings presented in this article. Section 3 describes the data collection applied in the study. The outcomes of the various analyses are outlined in Section 4, while Section 5 presents a discussion of the outcomes and draws major conclusions from the study.

## 2. Important factors differentiating between downloads and citations

This section lists ten important factors that may lead to divergence among download and citation counts and among rankings based on these parameters. These factors are also relevant when actually using and interpreting download and citation counts. The degree to which these factors cause bias in the outcomes of analytical usage studies depends not only upon the type of usage metric applied, but also upon the objectives of the analysis and the research questions addressed therein.

1. *Usage leak*. Not all full text uses of a publisher archive's documents may be recorded in the archive's log files. For instance, authors may download documents from an archive and share them with colleagues; "author copies" of submitted or published manuscripts in subscription-based journals and "Open Access" articles may be freely available on many locations on the internet. The use of such documents remains invisible in the log files of the publisher's archive.

2. *Citation leak*. Not all relevant sources of citations may be covered by the database in which citations are counted (citation leak). Citation analysis is mostly conducted in large, multidisciplinary databases covering mainly (though not exclusively) journals. Fields in which books or conference proceedings are important outlets may not be represented well in such databases. Important national journals in large, non-English speaking countries with a big internal scientific information market may not all be covered. For instance, comparing citation and download counts, Guerrero-Bote & Moya (2014) found evidence of a citation leak for journals publishing articles in non-English language.

3. *Downloading the full text of a document does not necessarily mean that it is fully read.* It is plausible to assume that a downloaded document has received the user's attention (except perhaps in the case of bulk downloading or data manipulation, see below), but this can also be in the form of reading the abstract and/or quickly browsing through the full text.

4. *Reading and citing populations may be different*. The user (reader) population and the author (citer) population may not fully coincide. Nicholas et al. (2005) distinguished three main user categories: practitioners, researchers and undergraduates. Kurtz & Bollen (2010) added a fourth category: the interested public. Of these four categories, researchers tend to publish papers – and cite other articles – in the scientific literature, but members from the other three categories tend to publish less research articles or no articles at all. Generally speaking, different sets of users can show substantially different usage behaviors even when they access the same documents (Kurtz & Bollen, 2010, p. 20).

5. *Number of downloads depends upon type of document*. The type of document or, more generally, a document's content is a crucial factor, even in the set of published journal manuscripts (Schloegl & Gorraiz, 2010; Paiva, et al., 2012). Editorials and news items may be heavily downloaded but poorly cited compared to full length articles. The same is true for un-refereed conference abstracts (Kurtz & Bollen, 2010, p.21). Kurtz et al. (2005b) highlight a document presenting an extensive review of the past astronomical literature that is heavily downloaded but poorly cited. Section 4 of the current article provides downloads-per-citation ratios for four main document types.

6. *Downloads and citations show different obsolescence functions*. Download and citation counts both vary over time, but in a different manner, showing different maturing and decline rates. Contrary to citation data, usage information is available in near real time. During the first few months after online publication, documents may be heavily downloaded but hardly cited at all, but after 4 years the number of downloads would substantially decline whereas the number of citations reach their peak value. As a result, rankings of a given set of articles according to usage or citation counts up-to-date may vary over time as well. Section 4 presents more data on obsolescence.

7. *Downloads and citations measure different aspects*. Short term downloads tend to measure readers' awareness or attention for document – Kurtz & Bollen (2010) characterize users as "current awareness checkers" –, whereas citations result from authors' reflection upon the literature used in the research process, leading to a selection of what the authors perceive as the most significant ones. Moreover, Kurtz et al. (2005b) found that the historical component in the usage obsolescence function has no counterpart in the citations, and hypothesized that this is due to the fact that many articles are downloaded for their historical interest, but do not directly influence current research problems. Perhaps the term "background reading" can be used in this context.

8. *Downloads and citations may influence one another in multiple ways*. In order to be cited, articles tend to be read and, hence, downloaded first. In this sense downloads lead to citations. But the reverse is true as well. Articles may gain attention and be downloaded when they are cited; in this sense citations may lead to downloads. But the time delays with which the assumed effects are visible are different. (e.g., Kurtz et al., 2005b; Moed, 2005; Kurtz & Bollen, 2010;). Kurtz et al. (2005b) even include what they term "learner's or student's use" as a separate factor in their usage model.

9. *Download counts are more sensitive to manipulation*. While citations tend to be regulated by the peer review process (and author self- citations can be easily detected), download counts are more sensitive to manipulation. Individuals may download their own papers numerous times or instruct for instance their students to do so. Downloading of complete journal issues or (annual) volumes in one single user session in order to produce one's own print version of journals should be distinguished from manipulative behavior, but may affect download counts as well.

10. *Citations are public, usage is private*. While citations in research articles in the open, peer reviewed literature are public acts, downloading documents from publication archives is essentially a private act. Use and publication of usage data involves privacy issues when aggregated at the level of individuals, institutions and providers (Kurtz & Bollen, 2010). Usage data do not only relate to sheer counts of downloads from publication archives, but also to contextual information on other documents downloaded by the same user, or to sharing or annotating of documents in reference managers.

# 3. Data collection

Data on numbers of full text article downloads was collected from two perspectives: that of the *downloaded* documents and that of the *downloading* users. In the analysis of downloaded documents, numbers of article downloads and citations were collected in two sets, one at the level of journals and the second at the level of individual documents.

1. *Journal Level Data*: the first set of data contained downloads data for all around 1,800 journals covered in ScienceDirect™, Elsevier full text database. This dataset is indicated as the "Total Set" throughout this article. In addition, it contained citation data to these journals extracted from Scopus™, Elsevier's database on research articles in about 20,000 journals published by 5,000 scientific-scholarly publishers. Both download and citation data relate to the time period 2004-2010, and were aggregated by year and by journal.

2. *Document Level Data*: Download and citation counts on a *per document basis* were collected for all documents published in 62 ScienceDirect™ journals between 2008 and 2012 covering all domains of science and scholarship. For a full list see ANNEX A1. Downloads and citations counts on document level are up to September 2013. This set is labelled as the "62 Journal Set" throughout this article.

In the analysis of downloading users, data on the number of full text data was collected in three sets:

1. *At the level of user sessions*: Data related to downloads from ScienceDirect made by two European academic institutions during 2002-2003.

2. *At the level of user institutions*: Data on downloads from ScienceDirect of documents published during January 2009-May 2013 made from selected institutions during the same time period.

3. *At the level of user countries*: Data on downloads from ScienceDirect of documents published during January 2009-May 2013 made from selected countries (China and UK) during the same time period.

It must be noted that the journals studied are *not* a random sample from the set of journals in ScienceDirect. On the contrary, the aim of the selection was to include journals from different disciplines and cover all major disciplines, in order to study differences among disciplines, and also to include journals that were originally sections of one and the same "parent" journal, so that one could even obtain indications of differences within a journal.

## 4. Results

*Downloads vs. citations of an individual article*

The aim of this section is to provide basic information on full text downloads, and to show a common longitudinal pattern of download and citation counts for an individual article.

Figure 1 presents for one particular article the number of downloads shown on the left vertical axis and the number of citations shown on the vertical right axis over each month after publication. The article is taken from the Journal of Wind Engineering and Industrial Aerodynamics. Downloading of an article from ScienceDirect is technically possible when the final version of the manuscript corrected by the authors is made available online. The date at which this occurs is the online publication date. It is important to note the different phases of publication i.e. corrected proof and corrected paginated proof as they are seen to generate different downloads patterns. As can be expected, the corrected proof which became available in March 2008 generated over 60 downloads followed by over 150 downloads when it was paginated in August 2008. At this date, the journal issue in which an article appears is complete, all its documents are online, and the downloading of its articles boosts in the early periods in the article's age which generate the highest downloads figures in an article's life cycle. It could be explained as current awareness activity when readers keeping abreast in their area of research and closely reading content as it becomes available**.** Variations between the 10th and 40th month are probably due to seasonal influences and academic life cycles. Figure 3 illustrates the effect of these influences upon downloads from a number of countries.

At this point, no citations are recorded which is expected for a 4 months old article. The first citations of this article appear in the following year, approximately July 2009. As will be shown below, the rapidity at which citations occur depends upon the journal and its subject field. These citations can be assumed to be at least partially a result of the article's early heavy downloads followed by steady downloads rate in the months followed. However, it is also important to observe how citations might affect the article's downloads on the periods following the appearance of the first citation. Figure 1 shows that downloads rate is increasing in close proximity to the first citation during months 37-40 followed by an additional peak appearing after citations are recorded in months 40-43. These download peaks may be the result of citations, as the latter increase an article's visibility. Although earlier papers (e.g., Moed, 2005) provide evidence that citations may have a positive effect upon downloads, causality in the relationships between downloads and citations are not further investigated in the current article.

The pattern shown in Figure 1 is a common pattern that can be observed for the overwhelming part of documents analysed in the current article: full text downloading starts when the corrected proof is online; next, usage increases strongly when the article is paginated, followed by a rapid decline – although it should be noted that the time period between the article's online publication date  and the decline phase of its full text downloads varies across journals and disciplines (see Figure 5) and document types (see Figure 6); next, influences of seasonal and academic cycles are visible in the decline period; and finally the monthly number of downloads shows a revival when the article is cited. It should be noted, however, that the time period between an article's online publication date and its first received citation depends upon the journal, discipline and document type as well, and may be much shorter than that of the article represented in Figure 1.

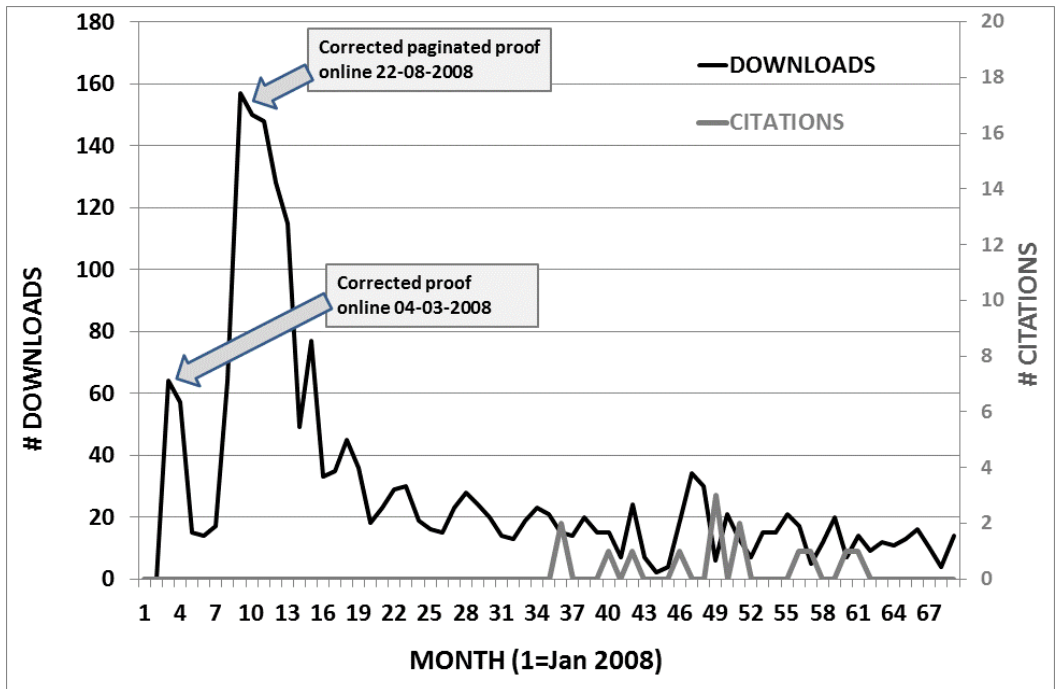

Figure 1. Longitudinal download and citation counts for an individual article

*Types of user session*

This section illustrates how full text downloads can be disaggregated by user session, and how normal user sessions can be distinguished from extraordinary ones.

The data presented in this sub-section were collected in an earlier study conducted by the first author of the current paper in 2003 (Moed, 2003). It aims to illustrate that one can roughly distinguish three types of user sessions, denoted below as "normal", "aggregate-normal" and "bulk". Data relate to downloads made from two European academic institutions during 2002-2003. In order to define user sessions from the usage log files, all downloads were arranged by IP address and by date and time of use. Next, a time out period was defined. The first download made from a particular IP address on a particular day marks the beginning of the first user sessions conducted from that IP address during that day. Next, if the time period between two subsequent downloads made from that IP address in that day exceeded the time out period, the current session was assumed to be ended, and a new session to start. As a time out period 30 minutes was chosen. In this way the number of sessions was not sensitive to small changes in the time out interval.

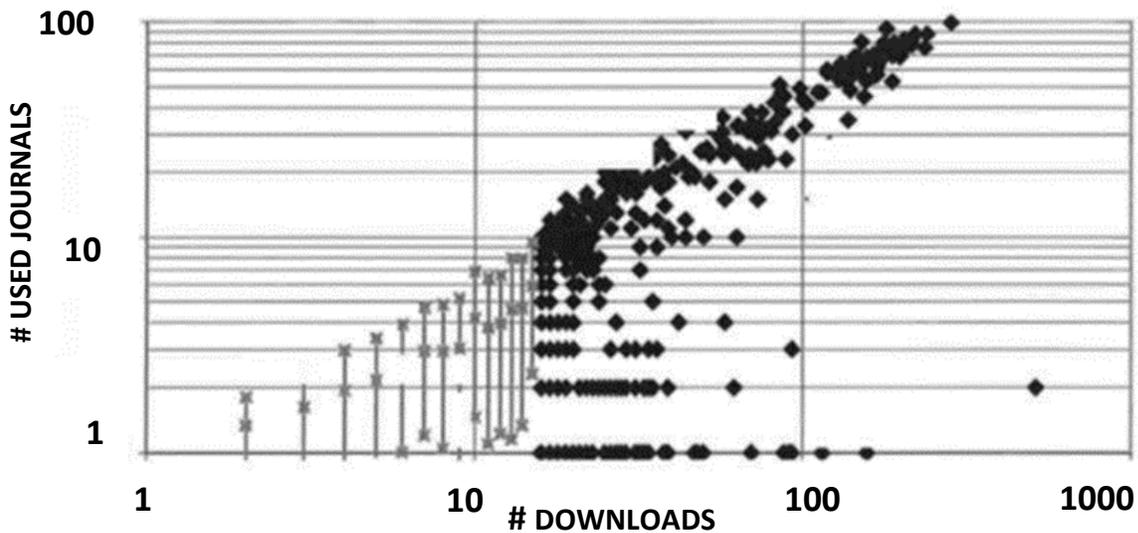

Figure 2. Distribution of number of full text downloads used journals by user session

Figure 2 analyses the number of journals used in a session. Black rhombuses indicate user sessions with more than 15 full text article downloads. The grey crossed squares and lines indicate the mean and standard deviation of smaller sessions with between 2 and 15 downloads. Focusing on sessions with more than 15 downloads, Figure 2 shows a cluster of sessions for which the number of used journal more or less proportionally increases with the number of documents downloaded in the session. These sessions are represented by rhombuses along the diagonal; these are probably aggregates of smaller sessions. The bottom of the graph shows a series of rhombuses all located on the horizontal axis with ordinate 1. These represent sessions in which documents from only one single journal were retrieved. Such sessions are assumed to represent "bulk" downloads. Figure 2 suggests that it is useful to distinguish three types of user sessions: "normal", "aggregate normal" – i.e., large number of downloads, but average mean time between two downloads and average downloads per journal, presumably resulting from downloads from a series of different users – and 'bulk" sessions, in which the number of downloads is large, the mean time between two downloads short, and the average number of downloads per journal relatively high.

*Downloads by user country and user institution*

The aim of this section is to show how seasonal influences and academic cycles and other factors may affect monthly full text downloads made by users from particular institutions or countries.

Figure 3 presents data on monthly full text downloads from ScienceDirect that users from 2 countries, China and UK, made between January 2009 and May 2013. The vertical axis gives the percentage of downloads in a month, relative to the county's sum of downloads during the total time period considered. If the number of a country's downloads would be constant over time, the monthly score would be around 2 per cent. Figure 3 clearly reveals seasonal and academic cycles. The countries show a regular pattern that is repeated each year. China's dip in the month of February is due to the celebration of Chinese New Year. In the UK, the months with the largest download activity are clearly March and November. Most if not all countries show seasonal influences and/or academic cycles in their aggregate download counts per month.

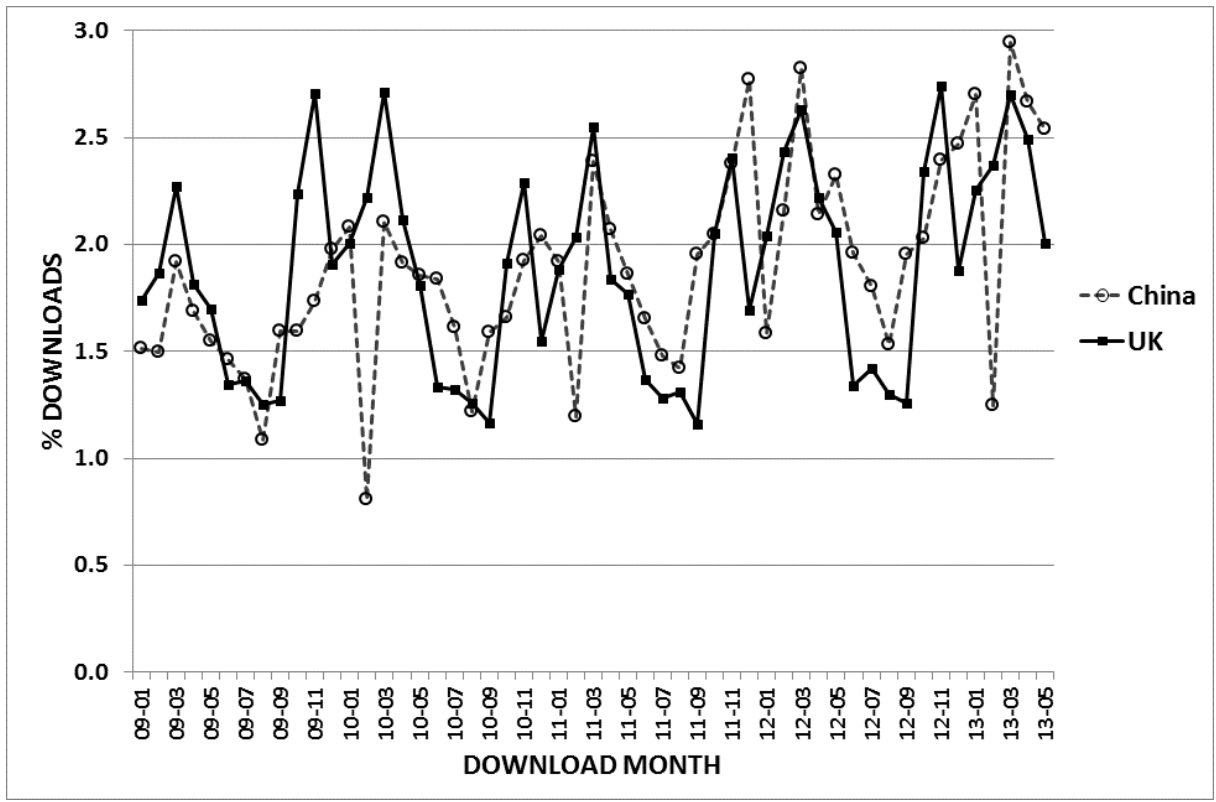

Figure.3. Longitudinal download counts for users from 2 countries: China and UK. The vertical axis gives the percentage of downloads in a particular month, relative to the country's sum of downloads during the total time period considered.

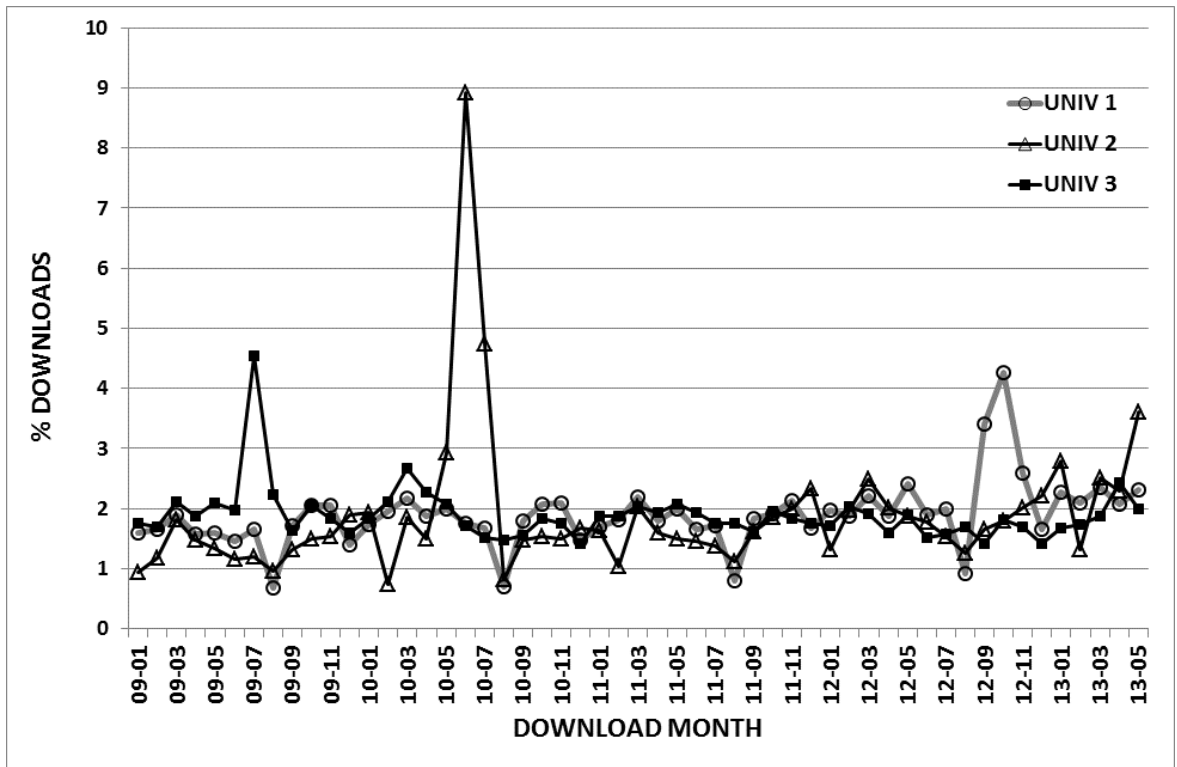

Figure 4. Longitudinal download counts for three user institutions. The vertical axis gives the percentage of downloads in a month, relative to an institution's sum of downloads during the total time period. For University 2 the actual percentage of downloads in July 2010 is 9 %, which is 4.5 times the level one would find if the number of an institution's downloads would be constant over time.

Figure 4 shows the percentage of downloads per month from three distinct institutions, relative to the total number of downloads an institution made during the entire time period analysed. It illustrates that large variations may exist in the number of downloads made by members of one single institution, variations that cannot be ascribed to seasonal influences or academic cycles.

University 1 represented in Figure 4 participated in a national research assessment exercise, in which research staff members submitted full text PDF downloads of their best articles to an evaluation agency for assessment by an expert panel, with a submission deadline in October 2012. For the peaks of Institutions 2 and 3 no explanation is available as of yet. Whether or not these peaks are caused by bulk downloading can be examined by grouping the downloaded articles by user session and by journal, and determining the number of downloads per journal in a session, as illustrated in Figure 4.

In order to give an estimate of the frequency at which such peaky behaviour revealed in Figure 4 occurs, and of the effect it has on the total number of downloads, an analysis was made of all about 7,000 institutions making downloads in each month. After calculating the first and third quartile of the distribution of an institution's monthly scores, outlier months were defined as months in which the number of downloads exceeds the value *Q3+k\*(Q3-Q1),* in which *Q1* and *Q3* represent the first and third quartile, respectively, and *k* is a positive constant. If a particular outlier month was identified, the "surplus" in that month was defined as the difference between the actual number of downloads and the average number of downloads per month calculated over all months except those showing an outlier. Setting the value of *k* in the above definition of outlier to 2.0, it was found that 45 per cent of institutions revealed an outlier score in at least one month, and that the total surplus value due to outliers across all institutions amounts to about 4 per cent of the total number of downloads made by all institutions (including the ones not showing any outlier) during the entire time period. Setting the value of *k* to 3.0, it was found that 27 per cent of institutions account to an overall surplus of 3 per cent.

*Obsolescence patterns per journal and document type*

This section shows how monthly download counts of articles vary over time, and how longitudinal patterns differ among types of document, journals and disciplines.

Figure 5 shows the average number of downloads per full length article over time for 6 journals covering social, applied, life, clinical medicine, mathematics and humanities, respectively. The overall phenomenon seen in Figure 5 is that all journals display peak downloads in the first months following publications, despite the difference in the amount of downloads which varies considerably between journals. Yet, there are differences among the represented journals in the month in which download counts peak. For instance, for the journals in clinical medicine and life sciences downloads peak *one* month after the month in which they were published online, whereas for the applied science and the mathematics journal in the *seventh* month. Moreover, large differences exist in the decline rates in the various journals. These decline rates themselves tend to decline as the documents grow older. This is consistent with the two-factor models explored by Parker (1982) and adopted by Moed (2005), and the four-factor models explored by Kurtz et al. (2005b).

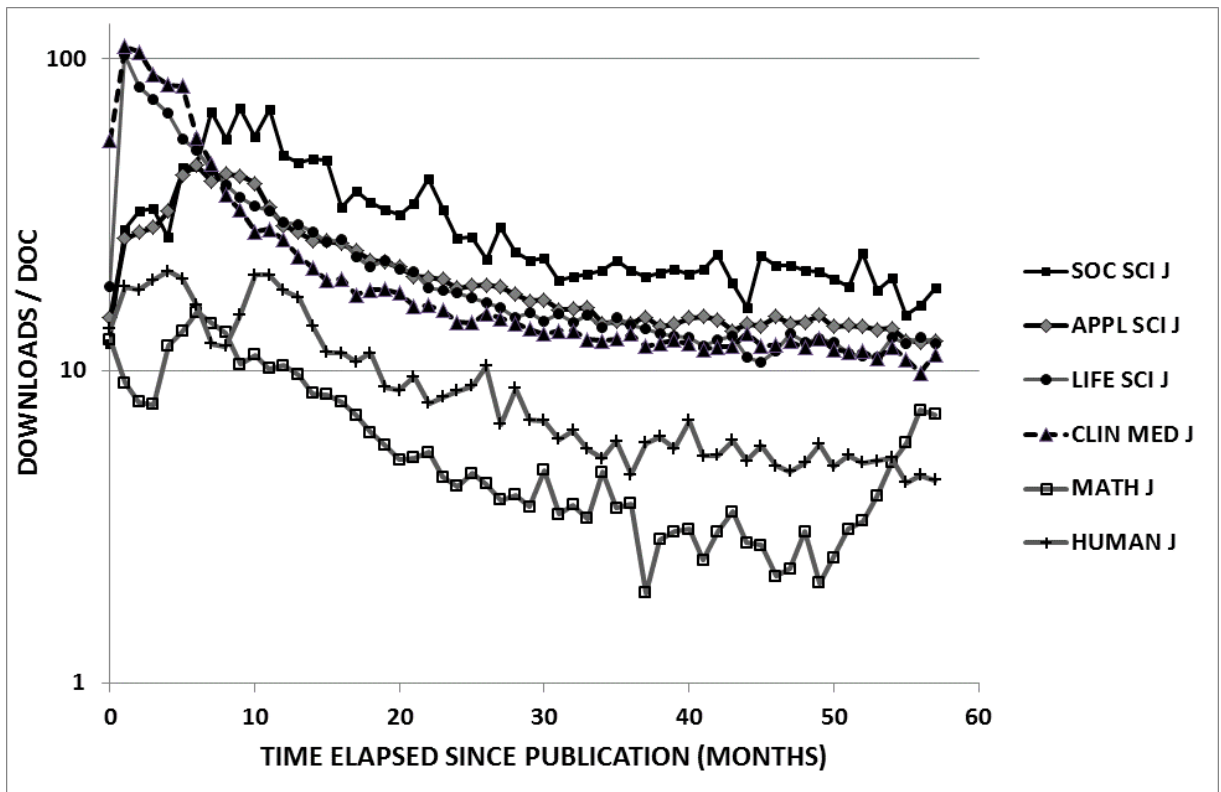

Figure 5. The number of downloads per full length article as a function of the articles' age for 6 journals covering the subject fields of social sciences (SOC SCI), applied sciences (APPL SCI), life sciences (LIFE SCI), clinical medicine (CLIN MED), mathematics (MATH) and humanities (HUMAN), respectively. Elapsed time 0 indicates the month in which the articles were published.

Figure 6 displays the development of downloads over time for four document types in the set of 62 journals: full length articles (Full text article (FLA), reviews (REV), short communications (SCO) and editorials (EDI). As can be seen in the graph, reviews, short communications and editorials reach their peak downloads in the first month after publication, and full length articles in the third month. Short communications and editorials show the most rapid decline during the first few months after publication. After two years, the decline rates of the four types are similar. The level of downloads is highest for reviews, and lowest for editorials, at least in the set of 62 journals.

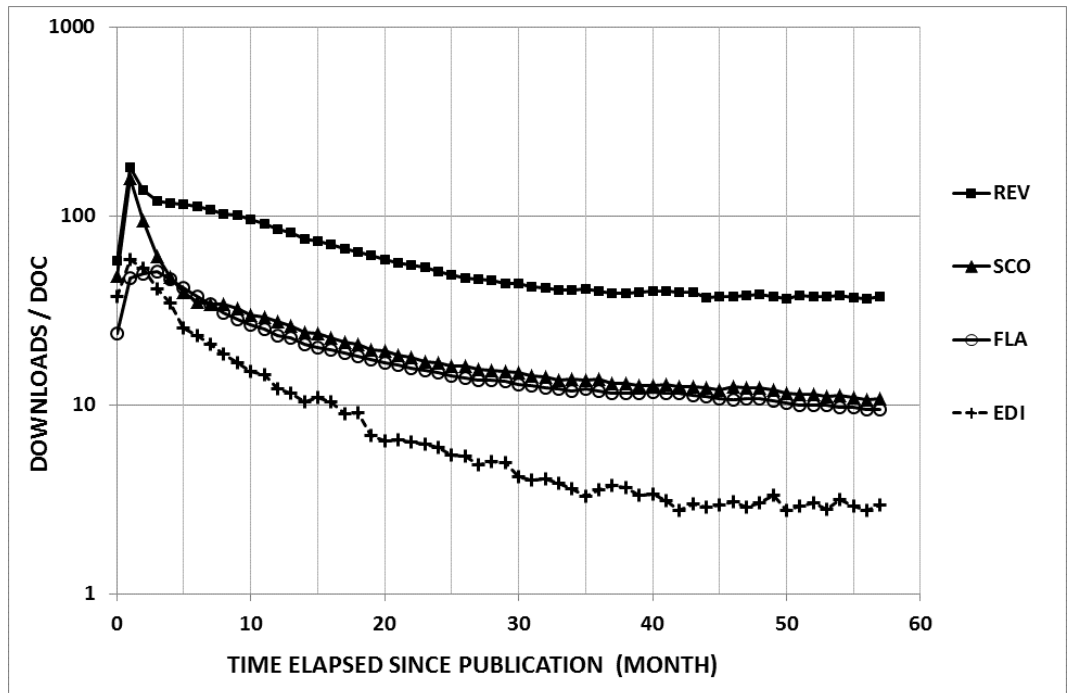

Figure 6: The number of downloads per document type as a function of the documents' age (62 Journal Set), for 4 document types published in the 62 journal set: full length articles (FLA), reviews (REV), short communications (SCO) and editorials (EDI). Elapsed time 0 indicates the month in which the articles were published.

*Download-versus-citation ratios*

This section provides insight into how frequently documents are downloaded compared to their citation rate and how this ratio of downloads per citations changes over time and varies across journals and disciplines.

Applying a diachronous approach, Figure 7 presents for documents published during 2008-2009 the ratio of the *accumulated* number of downloads and citations collected up until a particular month as a function of the documents' age in that month, or, in other words, as a function of the time elapsed since their online publication date, expressed in months. In this figure the documents from all journals in the 62 Journal Set are aggregated into one "super" journal. Ratios of downloads and citations are calculated for four types of documents: editorials, full length articles, short communications and reviews. Figure 7 clearly shows that the ratio of accumulated downloads and citations very much depends upon the type of document and upon the time elapsed since their publication date. For full length articles, reviews and short communications this ratio reaches a value of about 100 in the 45th month after online publication.

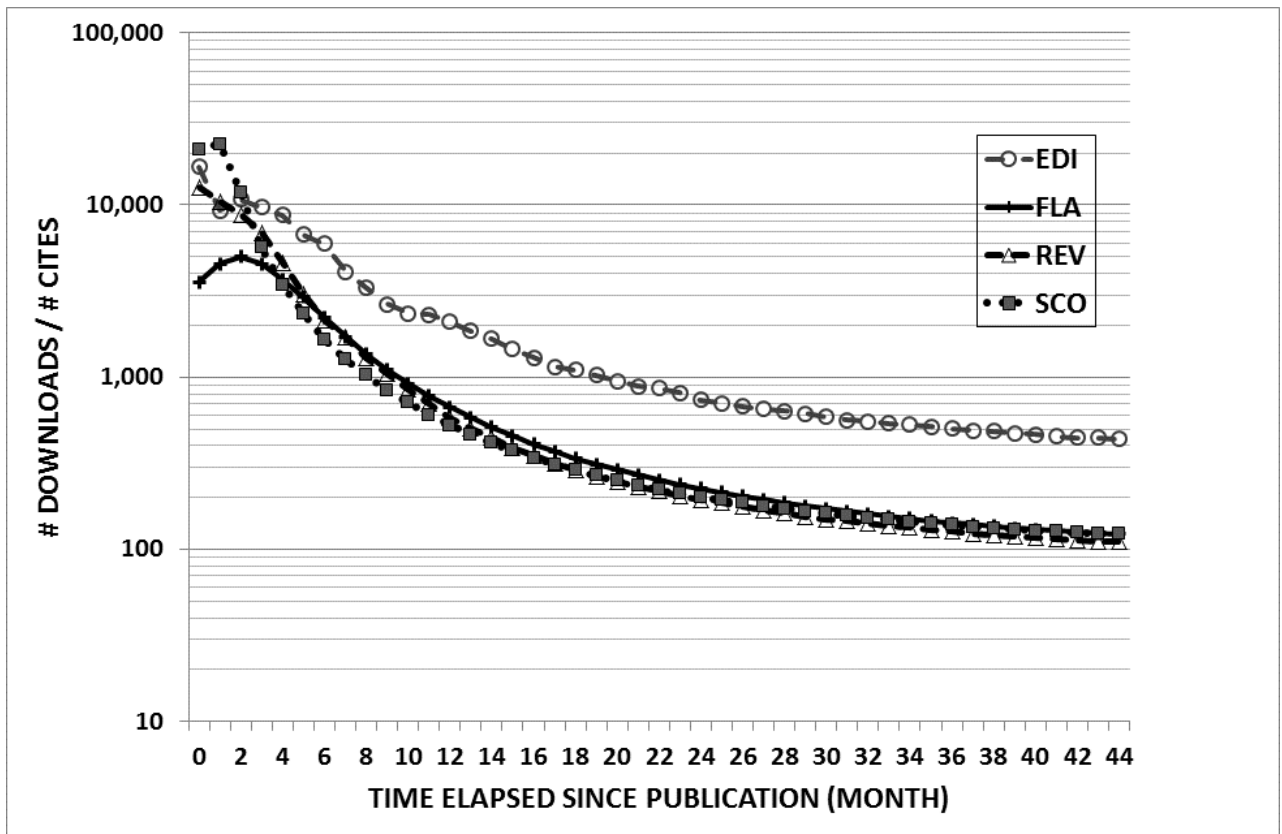

Figure 7: Ratio of accumulated downloads and citations of documents as a function of their age (62 Journal Set). EDI: Editorials; FLA: Full Length Article; REV: Review; SCO: Short Communications. Elapsed time 0 indicates the month in which the articles were published.

Figure 8, however, shows large differences in this ratio among the 62 journals. It displays on the vertical axis the ratio of accumulated downloads and citations for the aggregate of full length articles published in each of the journals in the 62 Journal Set, and on the horizontal axis the number of articles published in a journal during 2008-2009. Both downloads and citations were counted during the first 45 months after online publication date. Each symbol represents a particular journal. Distinct symbols indicate the main discipline covered by a journal. Figure 8 shows that journals in social sciences and humanities tend to have large ratios of downloads versus citations and several mathematics periodicals relatively low ratios. Clinical medicine journals show large variations. It must be noted that Figure 8 revealed that the ratio of accumulated downloads and citations changes with the length of the time period during which they are counted.

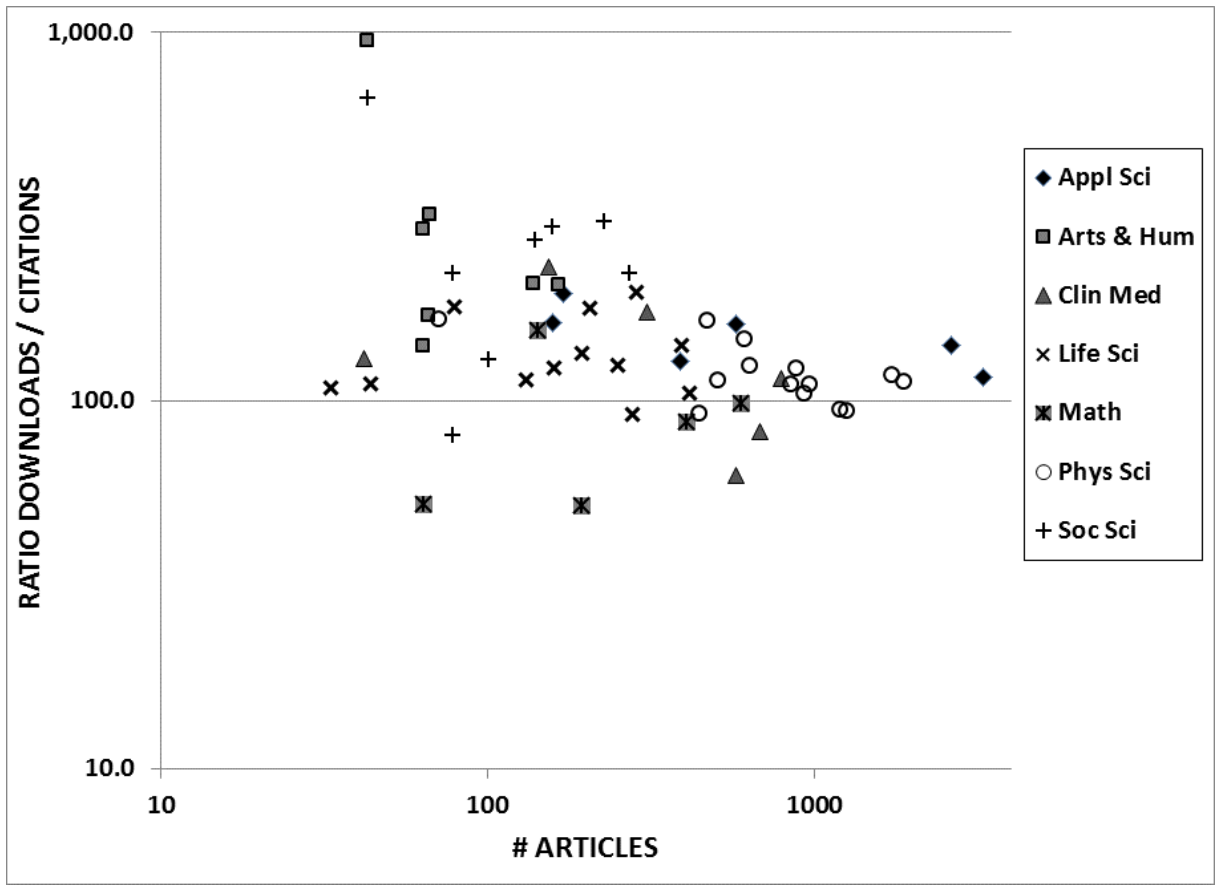

Figure 8. Ratio of accumulated downloads versus citations for full length articles in 62 journals

*The skewness of the article download and citation distribution*

This section compares the percentage of articles that are not cited with the percentage of articles that are not downloaded, and shows how these percentages change over time and vary across journals and disciplines.

Figure 9 displays the percentage of "unused" documents as a function of the time elapsed after their publication date. It relates to all documents published in 2008 and 2009 in journals in the 62 Journal Set. The curves labelled as "Citations" and "Downloads" give the percentage of documents that are uncited or not-downloaded in the month indicated on the horizontal axis. The lines labelled with the term "cumulative" give the percentage of documents that are unused during the entire time period from publication date up until and including the month indicated on the horizontal axis.

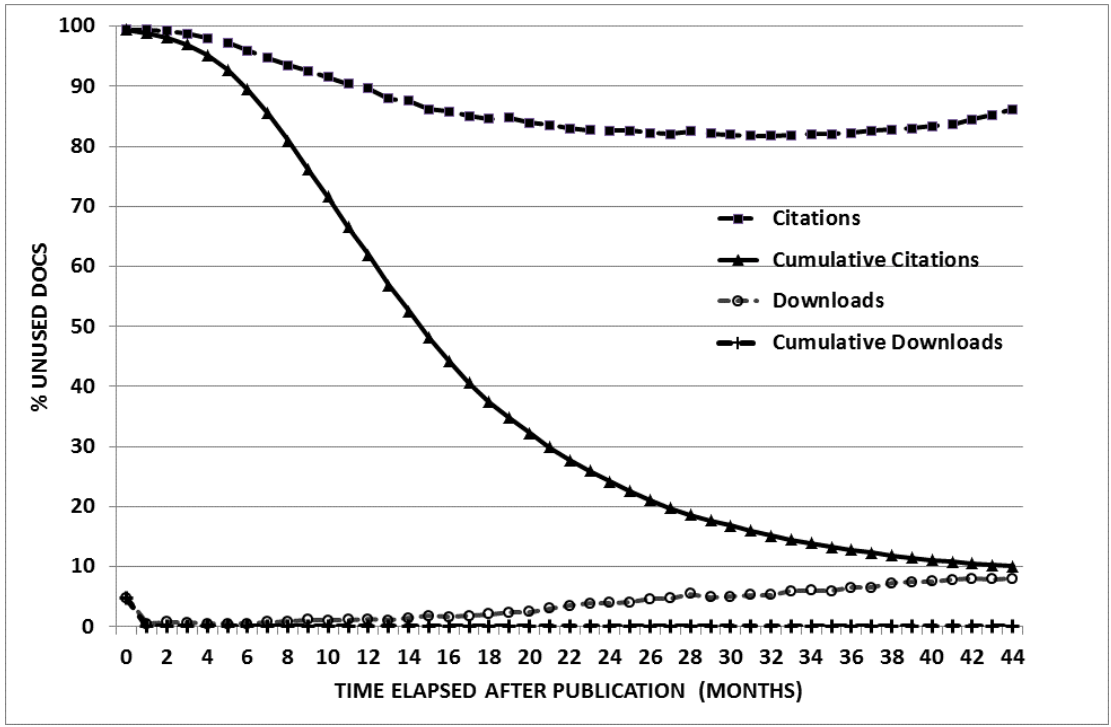

Figure 9. Percentage of unused documents as a function of their age (62 Journal Set, Full length articles. Elapsed time 0 indicates the online publication month.

Figure 9 shows that all documents in the set have been downloaded at least once during the first 45 months of their life cycle (including the online publication month). In fact, almost all documents are shown to be downloaded at least once in the 2nd, 3rd or 4th month after publication. In the 45th month after publication date, almost 10 per cent of documents are not downloaded anymore. By contrast, 87 per cent is not cited in that month. However, considering cumulative counts during the first 45 months after publication date, 10 per cent of documents is not cited. The degree of overlap between the (almost) 10 per cent of articles not downloaded in the last month and the 10 per cent of articles never cited up until that month was found to be about 30 per cent.

Figure 10 offers a different look at the skewness of the download and citation distributions, for one particular journal, *Topology and its Applications*. In a first step, articles published in 2008-2009 in the journal are sorted by descending number of downloads collected during the first 45 months since publication. Next, the functional relationship is calculated between the cumulative percentage of articles in the journal and the cumulative percentage of its downloads. The same is done for citations. The two functional relationships are plotted in Figure 10. It shows, for instance, that the top 50 per cent of articles in terms of citations accounts for 93 per cent of all citations to the journal, while the top 50 per cent of articles in terms of downloads accounts for only 64 per cent of all downloads. In other words, citations are much more skewedly distributed among the journal's articles then are downloads. The difference between the two percentages amounts to 29 per cent.

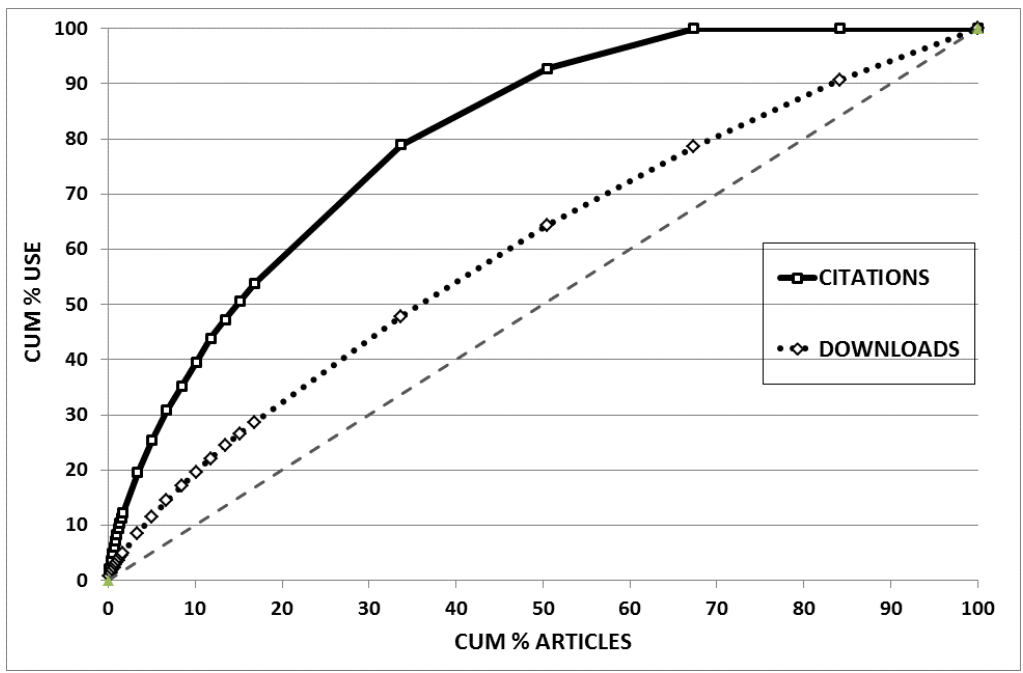

Figure 10. Cumulative percentage of downloads and citations as a function of the cumulative percentage of articles in the journal *Topology and its Applications*.

Figure 11 shows the distribution of these values for each journal in the 62 SET. All differences are positive, indicating that the download distribution is less skewed than the citation distribution. The mode of the distribution displayed in Figure 11 is 10-11. The 4 journals with the largest difference are: *Topology and its Applications*, *Differential Geometry and its Application*, *Annals of Pure and Applied Logic*, and *Lingua.* The two journals with the smallest difference are Cancer Letters and European Journal of Cancer.

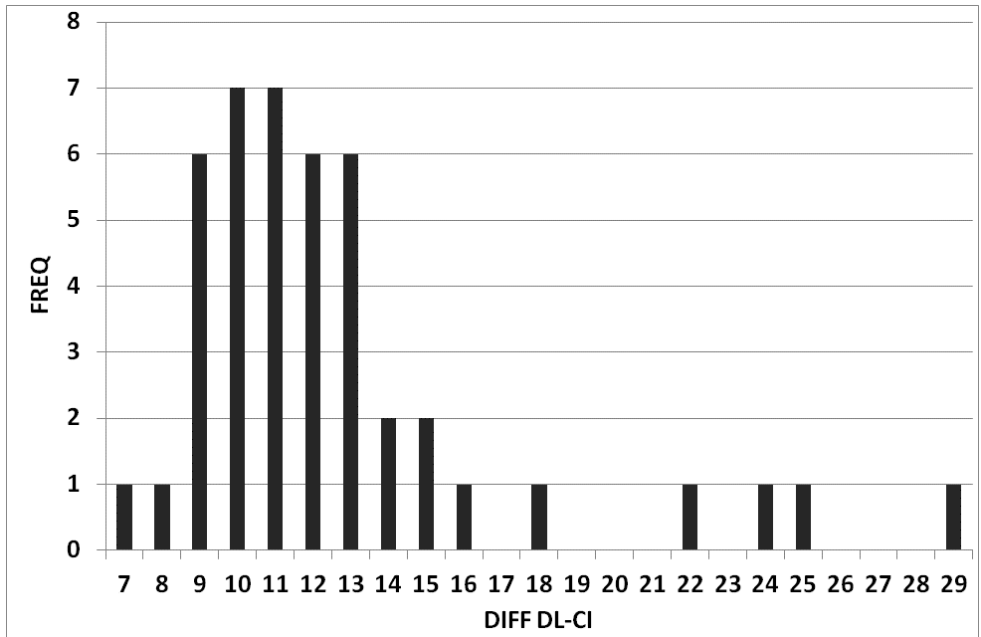

Figure 11. Distribution of journals (62 Set) according to the difference between the cumulative percentage of downloads and that of citations, accounted for by the 50 per cent most frequently downloaded or cited articles, respectively. (DIFF DL-CI).
.

*Statistical correlations between downloads and citations at the journal level*

In this section it is investigated whether the journals in a particular discipline that are highly cited are also the most often downloaded ones, and whether differences exist in this respect among disciplines.

For the set of all ScienceDirect journals we examined the rank correlation between downloads and citations at a journal level over time. Figure 12 shows how the Spearman rank correlation coefficient (Rho) between downloads and citation counts depend upon the year of citation and the year of download. The highest values of Rho are found between the number of downloads made during the publication year and the number of citations received in the second, third, and fourth year after publication year, with values between 0.6 and 0.7.

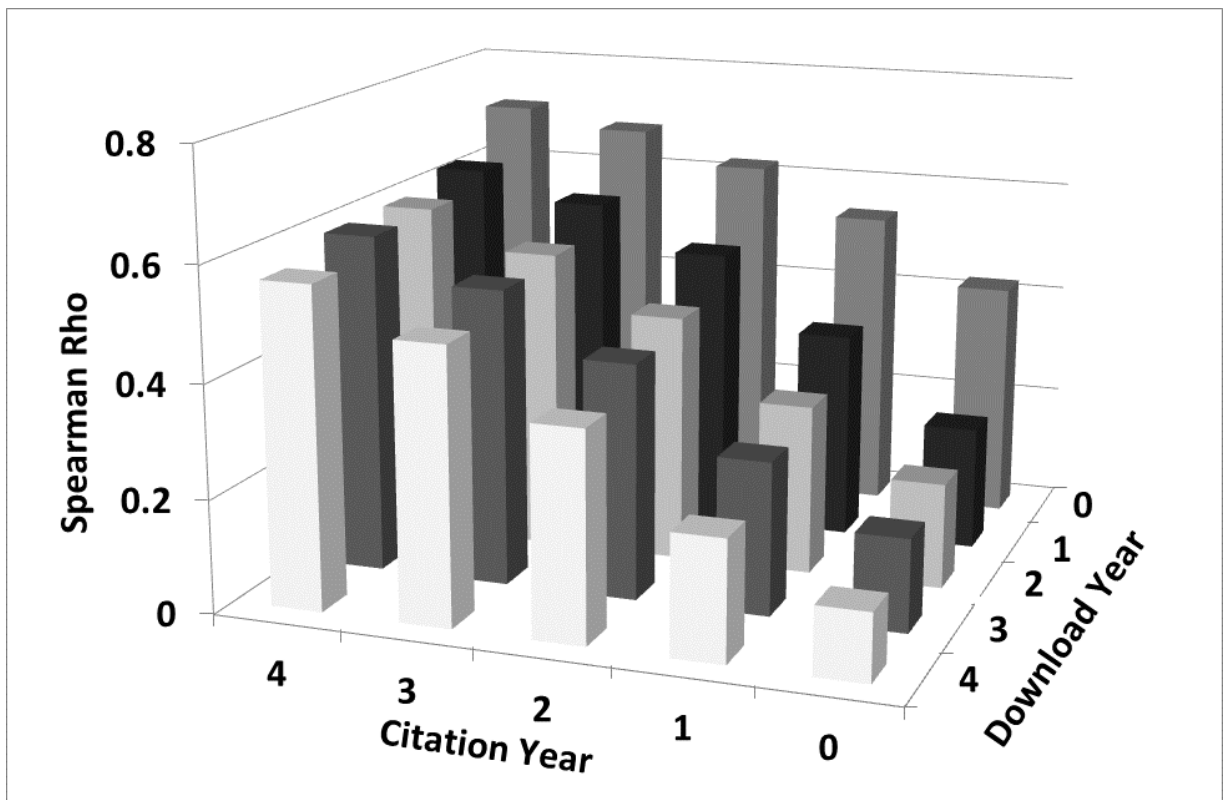

Figure 12. Rank correlation (Spearman's Rho) between downloads and citations as a function of citation and download year (Total Set). Citation or Download Year 0 indicates the publication year, year=1 one year after publication year, etc.

Figure 13 is based on download counts in the year of publication and citations in the second year after publication year (e.g., publication year 2005, downloads made in 2005, citations received in 2007). It shows the Spearman rank correlation (Rho) per discipline. Analysing the correlation at the level of a discipline between a journal's average number of downloads per article against the number of cites per article, Figure 13 shows that in the areas of *chemical engineering*, *biochemistry & molecular biology*, *neuroscience* and *veterinary sciences* downloads and citations are highly correlated. Disciplines which show the lowest rank correlation coefficients between downloads and citations are *arts & humanities*, *dentistry*, *health professions*, *psychology* and *social sciences*.

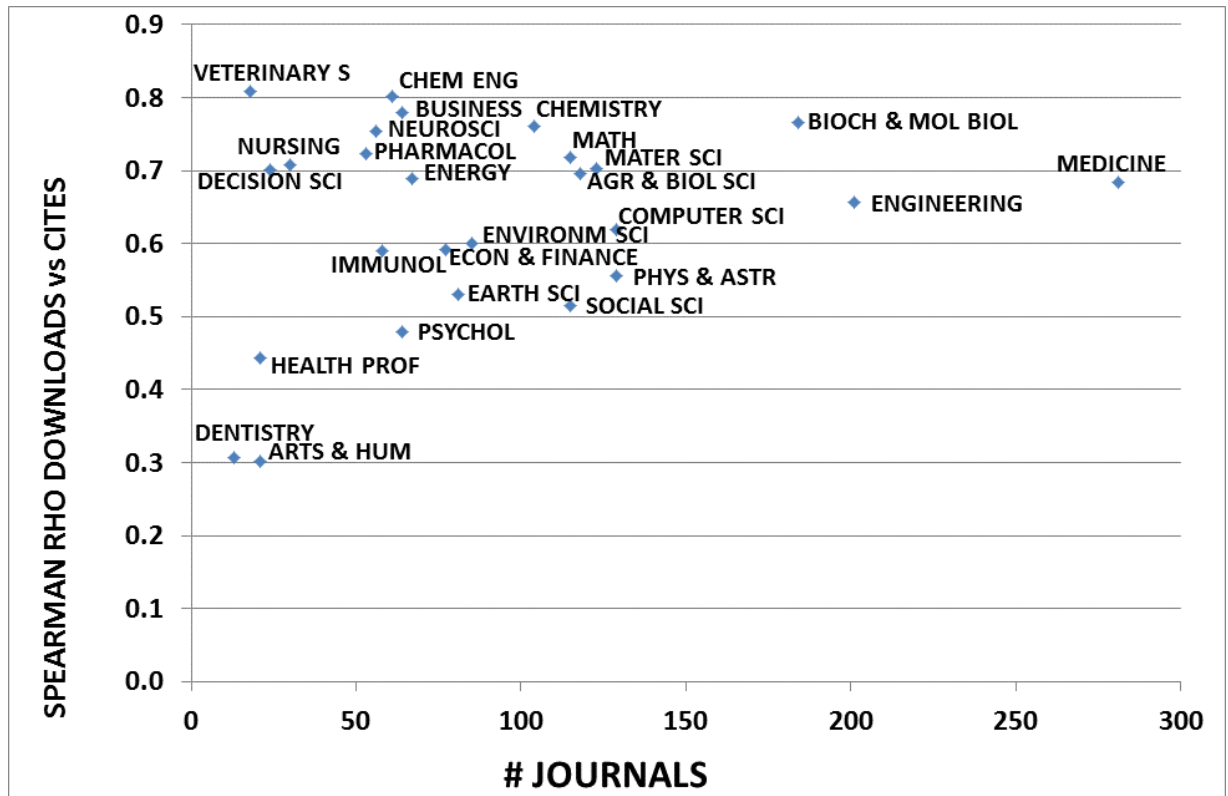

Figure 13. Correlation between downloads and citations at the journal level by discipline (Total Set).

*Statistical correlations between downloads and citations at the article level*

This section examines whether the articles in a particular journal that are highly cited are also those that are the most heavily downloaded and vice versa, and whether in this respect differences exist among journals and disciplines.

In all 62 journals the Spearman rank correlation coefficient between downloads and citations is positive. It ranges between 0.30 for *Medieval History* and 0.80 for *Biochimica et Biophysica Acta – Gene Regulatory Mechanisms*. Among the 10 journals with the lowest values, 2 are from arts & humanities, 3 from social sciences, 4 from mathematics, and one from applied sciences. In the set of 10 journals with the largest rank correlation, 8 cover life sciences, and 2 clinical medicine.

Figures 14a and 14b illustrate a case study showing how strongly the value of a linear (Pearson) correlation may depend - in statistical terms - upon outliers and –in editorial terms – upon document types, and why in this case rank correlation coefficients such as Spearman's Rho used in this article are more appropriate correlation measures.

The data relate to articles published in 2008-2009 and followed during a time period of 45 months. Both figures relate to the same science journal. Figure 14a includes 7 review articles, most of which are heavily downloaded and cited compared to normal articles. The Pearson correlation coefficient R amounts to 0.83 (which equals the square root of the value of $R^2$ in Figure 14a). This value is strongly determined by a few highly cited and downloaded documents which appear to be reviews. In fact, the Spearman rank correlation coefficient, based on ranks rather than absolute scores, of the scatter in Figure 14a is 0.65. Figure 14b shows that if reviews are omitted, the value of the Pearson correlation coefficient declines with about 25 per cent to 0.62, a value that is very similar to that of Spearman's Rho in Figure 14a. This outcome illustrates that the linear correlation coefficient (Pearson's R) may be strongly determined by a few outliers, and that in this case a correlation coefficient such as Spearman's Rho is more appropriate expression of the tendency towards statistical association then Pearson's R.

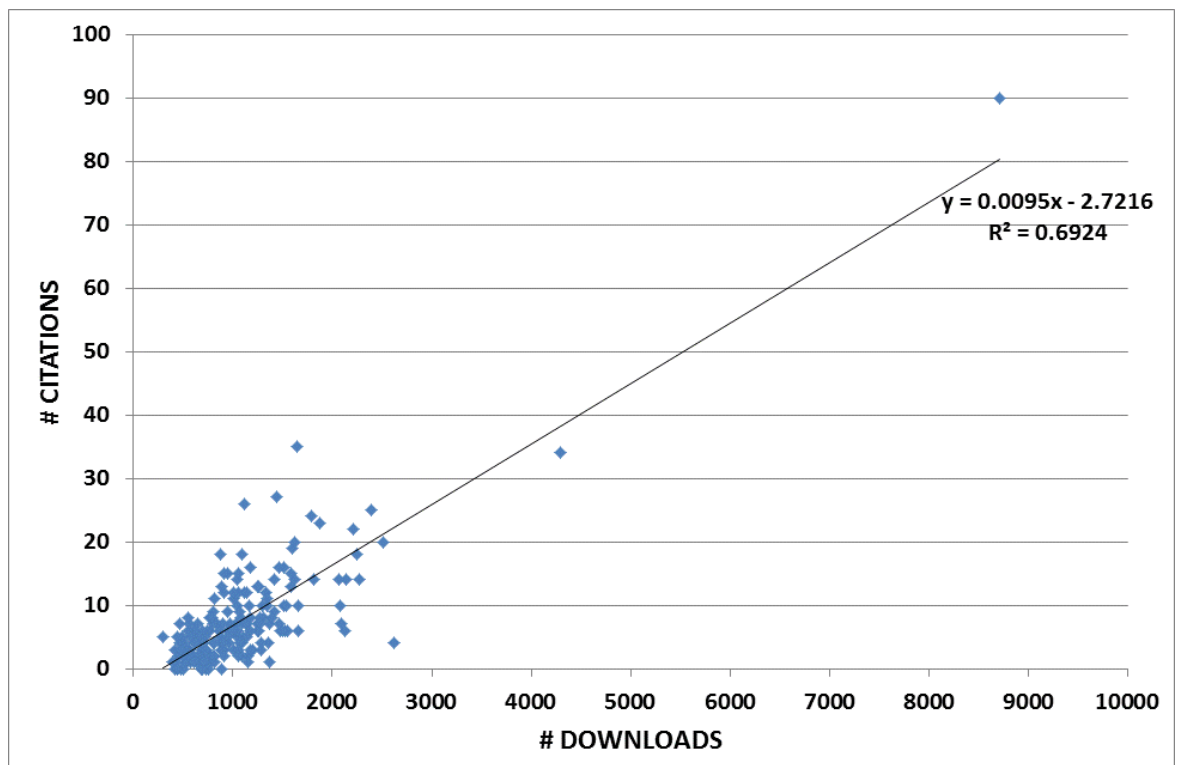

Figure 14a. Download versus citation counts for documents published in a science journal (all document types, including reviews)

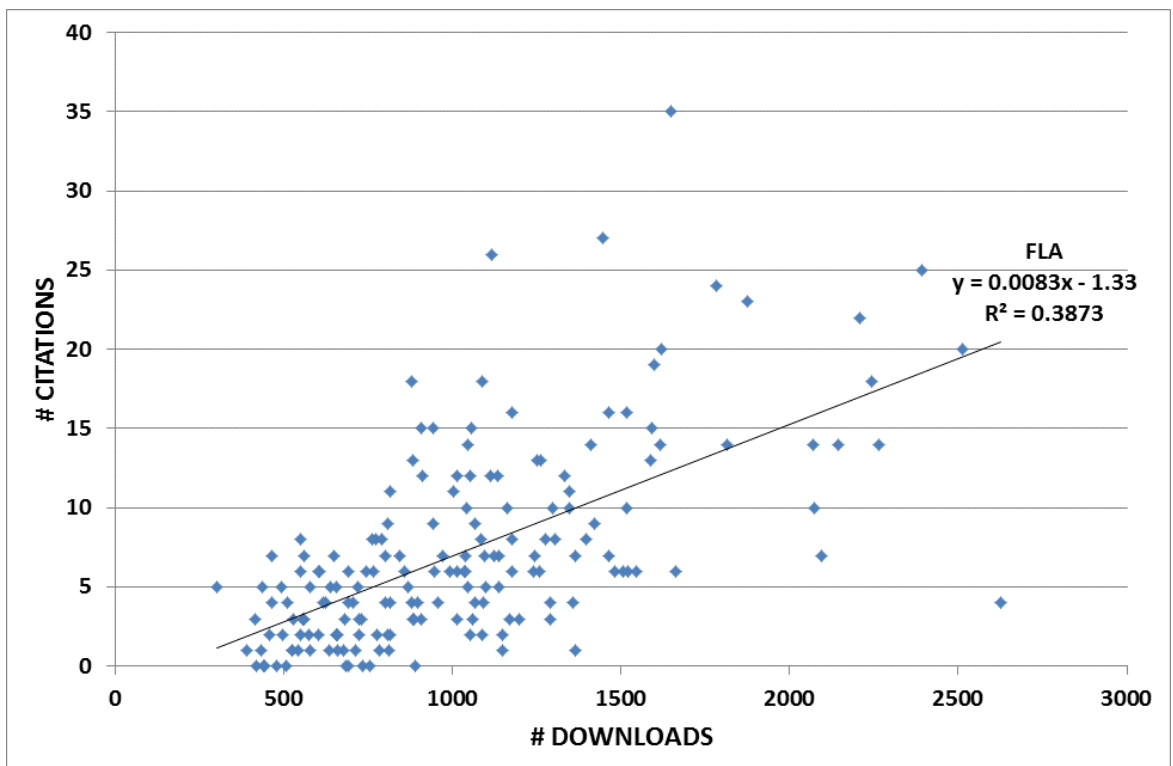

Figure 14b. Download versus citation counts for full length articles in the same journal as in Figure 15a. Reviews are not included in this graph

*Comparing correlations at the article level with those at the journal level*

This section investigates how the degree of correlation between downloads and citations at the journal level compares to the correlation between these two counts at the article level within a journal.

Figure 15 compares Spearman's Rho between downloads and citations *at the article level* in each of the 62 study journals (on the horizontal axis) with Spearman's Rho between the average download and citation rate *at the journal level* in the discipline(s) covered by a study journal (on the vertical axis). For instance, the Spearman rank correlation coefficient (Rho) between downloads and citations for articles in *Biochimica et Biophysica – Molecular Cell Research* amounts to 0.77. This journal is assigned to the discipline *biochemistry and molecular biology*; Spearman's Rho at the level of journals in this discipline amounts to 0.76. Figure 15 shows that the correlation coefficients at the article and journal level correlate positively themselves. Pearson's R and Spearman's Rho both amount to about 0.5.

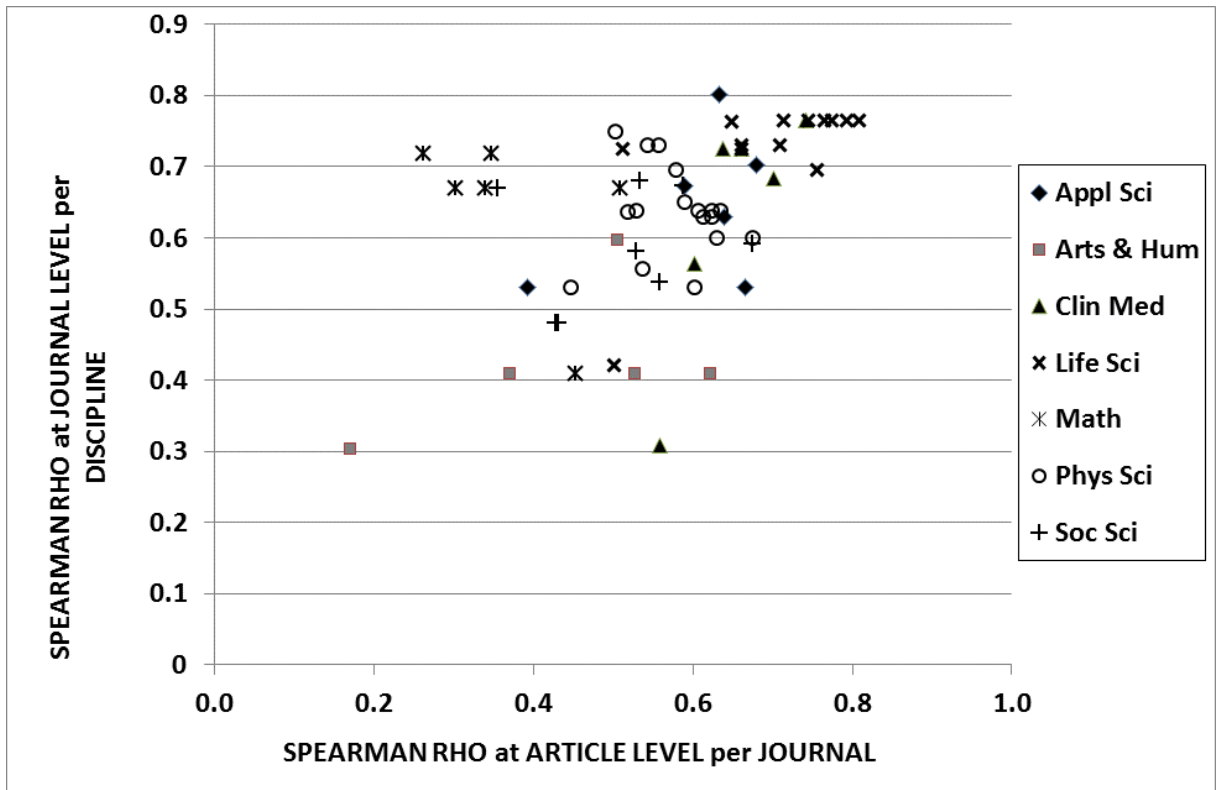

Figure 15. Spearman rank correlation coefficients between downloads and citations at the article level per study journal (62 Journal Set) and at the journal level per discipline covering a study journal.

*Overlap between sets of highly downloaded and highly cited articles*

This section examines whether the 10 most frequently downloaded articles are also the most heavily cited ones, and whether differences exist in this respect among journals and disciplines.

Figure 16 presents a scatterplot of downloads versus citation counts of articles in an applied science journal. The diagonal represents the linear regression line. It shows that the articles that are frequently downloaded (tentatively defined as those with more than 2,000 downloads) almost all have a minimum citation count of about 10. In other words, among the articles cited less than 10 times, there are no highly downloaded articles. This is so to speak one side of the correlation coin. But apart from this observation, the citation counts of the highly downloaded articles show a strong scatter. Such a scatter is even more clearly visible among the download counts for articles that are highly cited (tentatively, more than 20 times). But all these highly cited articles have a download rate that exceeds 500.

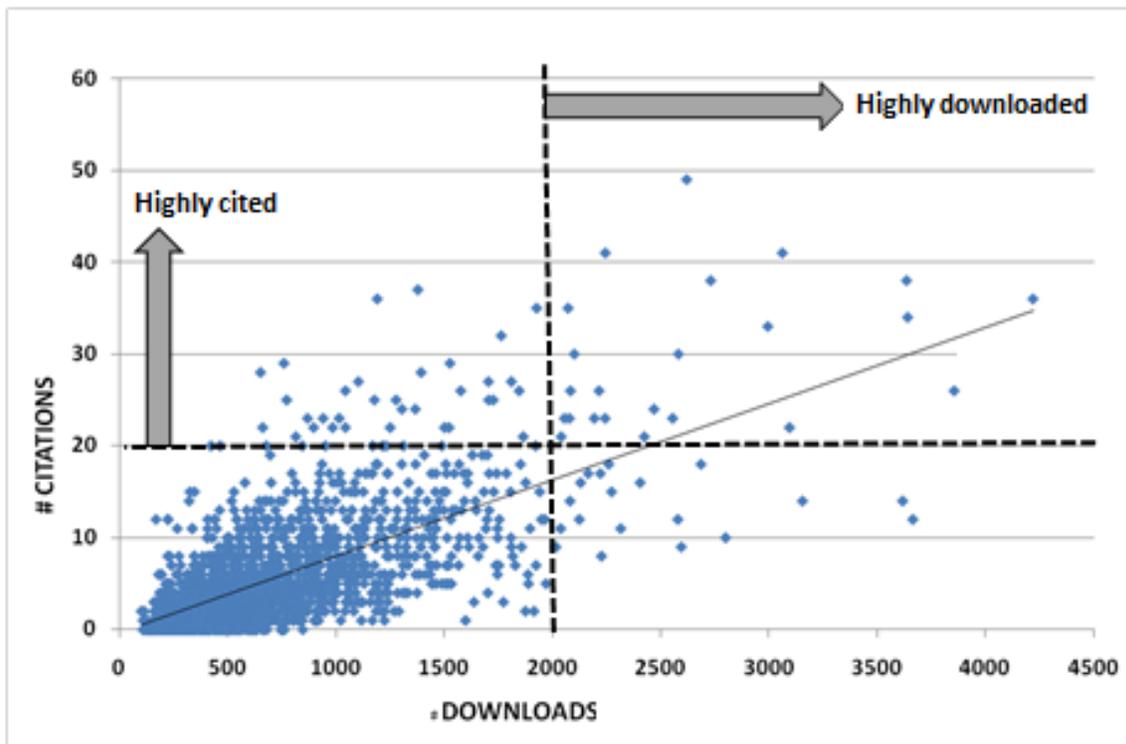

Figure 16. Downloads versus citation counts for a journal in applied sciences.

The vertical axis in Figure 17 indicates the number of articles that appear both in the top 10 most frequently downloaded articles *and* in the top ten most heavily cited ones. The figure shows that there is a rather strong positive correlation between Spearman's Rho for the association of downloads and citations at the document level within a journal on the one hand, and the degree of overlap among the top 10 sets in terms of downloads and citations in that journal on the other. This is in itself not surprising. Figure 17 gives an impression of what the implications of a certain degree of rank correlation between the two variables can be for the *top* of the rankings based on these variables. The two top 10 rankings for *Tectonophysics*, *Physica A*, *Tetrahedron Letters* and *Journal of Dentistry* have an overlap of one single article only. For top articles in *Biochimica et Biophysica Acta - Molecular Basis of Disease*, and *Stem Cell Research*, the overlap is 80 per cent.

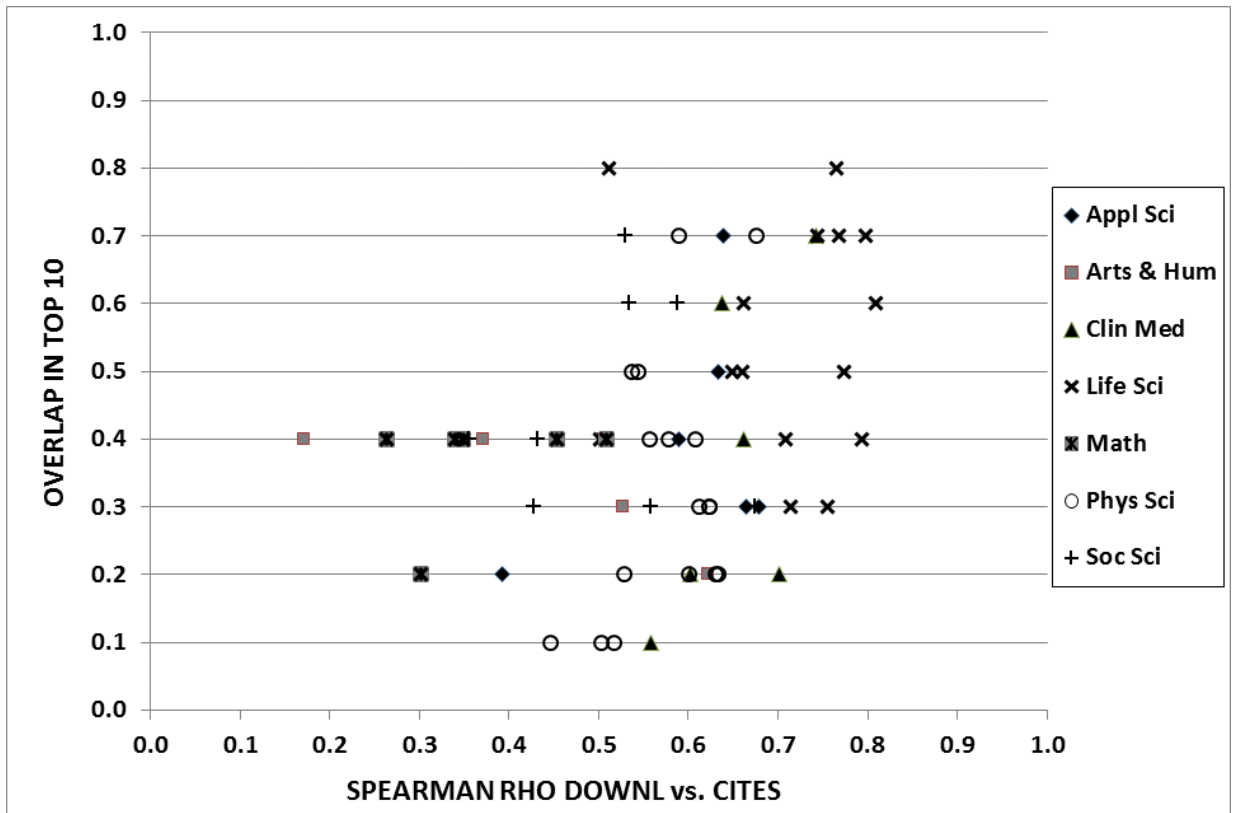

Figure 17. Spearman correlation coefficients between an article's download and citation counts by journal, and the degree of overlap between the top of the download and the top of the citation distribution.

*Comparison with findings presented by Kurtz et al. (2005b).*

Kurtz et al. (2005b) found evidence that "citations are a good predictor of downloads", but that "downloads are a poor predictor of citations". In this section their approach is applied to the data in the 62 Journal Set in order to examine whether the same conclusion can be drawn.

Table 1: Downloads versus citations using Kurtz et al. (2005b) data representation

| Citations | Downloads | | | | | | | | | | |
|---|---|---|---|---|---|---|---|---|---|---|---|
| | 16 | 32 | 64 | 128 | 256 | 512 | 1,024 | 2,048 | 4,096 | 9,092 | Total |
| 0 | 0.00 | 4.52 | 8.69 | 9.84 | 9.87 | 8.73 | 6.17 | 2.58 | 0.00 | ... | 2,797 |
| 1 | ... | 3.17 | 7.48 | 9.41 | 10.21 | 9.64 | 7.53 | 3.70 | 1.00 | 0.00 | 3,046 |
| 2 | ... | 2.32 | 7.32 | 9.49 | 10.86 | 10.98 | 9.19 | 6.00 | 2.58 | ... | 5,414 |
| 4 | ... | 1.00 | 5.58 | 8.33 | 10.57 | 11.74 | 10.73 | 7.58 | 4.86 | ... | 7,211 |
| 8 | ... | ... | 2.58 | 5.93 | 8.91 | 11.25 | 11.37 | 8.88 | 6.00 | ... | 6,127 |
| 16 | ... | ... | 0.00 | 2.32 | 5.61 | 8.88 | 10.42 | 9.22 | 6.00 | ... | 2,551 |
| 32 | ... | ... | ... | ... | 1.58 | 5.00 | 7.57 | 7.95 | 4.32 | 2.32 | 541 |
| 64 | ... | ... | ... | ... | ... | 0.00 | 2.81 | 5.00 | 1.58 | 2.00 | 64 |
| 128 | ... | ... | ... | ... | ... | ... | ... | 0.00 | ... | 1.58 | 7 |
| 256 | ... | ... | ... | ... | ... | ... | ... | ... | ... | ... | 0 |
| 512 | ... | ... | ... | ... | ... | ... | ... | ... | ... | ... | 0 |
| 1,024 | ... | ... | ... | ... | ... | ... | ... | ... | ... | 0.00 | 1 |
| Total | 1 | 39 | 808 | 2,702 | 6,027 | 9,602 | 6,757 | 1,620 | 189 | 14 | 27,759 |

*Legend to Table 1:* Table 1 shows for all articles published in 2008 and 2009 the relationship between accumulated download and citation counts collected during the first 45 month after online publication. The columns relate to the number of downloads and the rows to the number of citations. As in Tables 1 and 2 in Kurtz et al (2005b), data are binned in factors of 2. For example, the fifth column relating to downloads shows the number of citations for articles downloaded between 256 and 511 times. There were no articles in the study set downloaded less than 16 times. The actual numbers in the cells are the base 2 logarithm of the actual counts. Thus the number of articles that were downloaded between 256 and 511 times and that were cited between 4 and 7 times is 2 to the power 10.57 which equals 1,518.

Table 1 relates to all full length articles published during 2008-2009 in the 62 Journal Set, and to downloads and citations accumulated during the first 45 months after online publication. Its structure is identical to Tables 1 and 2 in article by Kurtz et al. (2005b) analysing downloads and citations recorded in 2000 in the NASA Astrophysics Data System. Looking *row-wise* at the number of citations (the left column in Table 1) and finding the most likely number of downloads for a given number of citations, the table shows a tendency that the most likely number of downloads increases with the number of citations, but not as strong as that found by Kurtz et al. (2005b) in their astrophysics database. In fact, for articles cited zero times the most likely number of downloads is between 256 and 512; the same is true for articles cited once. For articles cited 2-3 times and for those cited 4-7 times, it is 512-1023, and for papers cited between 8 and 15 or between 16 and 32 it is 1,024-2,047 downloads. Looking at the cells to the right and left of these maxima we see that the decline is on average around a factor of 2 (a difference of 1.09 in the base 2 log) for each cell, where each cell is a factor of 2 in number of downloads. But the standard deviation is large as well, obtaining in the base 2 log a value of 1.31.

Looking *column-wise* at the number of downloads and finding the most likely number of citations for a given number of downloads, Table 1 also shows a tendency that the most likely number of citations increases with the number of downloads. Looking at the cells above or below the maxima, the decline is 0.70 in the base 2 log, which is a bit lower than the decline rates obtained in the row-wise analysis presented above, but the standard deviation is only 0.42 which is much lower. Moreover, in the row-wise analysis focusing on the maximum number of downloads per citation bin, in most cases two subsequent citation bins revealed the same maximum number of downloads, while in the column-wise approach analysing the maximum number of citations per downloads bin, there is in all cases except one only one downloads bin with a particular maximum number of citations

.

From their analysis of astrophysics data, Kurtz et al. (2005b) concluded that "citations are a good predictor of downloads", but that "downloads are a poor predictor of citations". The results presented above for the 62 Journal Set do not allow for such a conclusion. There is perhaps even more evidence for the reverse conclusion, namely that downloads are a good predictor of citations and citations a poor or in any case a less valid predictor of downloads.

# 5. Discussion and conclusions

*The ratio of downloads and citations*

The main conclusion from the analysis at the level of ScienceDirect and Scopus as a whole is that during the first five years of the documents' lifetime the average download rate of documents in ScienceDirect is between 2 and 3 orders of magnitude (i.e., a factor of 100 to 1000) larger than the documents' citation rate. The ratio of downloads and citations declines with age of the used articles; after 5 years it is in the order of magnitude of 100. An analysis in the set of 62 study journals revealed that the rate of decline decreases over time, and that the value of the downloads per citation ratio seems to stabilize somewhat after three years or so. During the first few months after publication date the difference easily reaches a level of three orders of magnitude or higher. This is no surprise given the extremely low citation levels during these months.

 Section 4 revealed large differences in this ratio among disciplines, journals, and document types. Ratios tend to be higher in social sciences and humanities journals. Rather than reflecting an intensive usage behavior in these domains of scholarship, this outcome may primarily reveal a citation leak in the citation database, due to a limited coverage of sources in these fields. This hypothesis is consistent with the findings by Guerrero-Bote and Moya (2014) who analyzed the effect of the factor publication language upon downloading and citation behavior.

The observation in the set of 62 study journals that the ratio of downloads and citations is for editorials after 4 years almost four times that for full length articles or reviews indicates that editorials are not peer reviewed research articles and therefore tend to be less cited in original research contributions, but that they may attract interest for other reasons, for instance because they present an interesting personal opinion, an overview of the literature on a topic, or raise a more general issue in the research domain covered by a journal. This is consistent with observations made by Kurtz et al. (2005a, 2005b) and Kurtz & Bollen (2010).

*Longitudinal download patterns per user session, institution and country*

The fact that seasonal and academic cycles are reflected in longitudinal download patterns is not surprising. What is of interest is the peaky behavior at the level of user institutions – especially the appearance of disproportionally large numbers in particular months – and the apparent lack in many cases of solid explanations for such behavior. Even if the overall contribution of the number of downloads made in peak months across institutions is only a few per cent of the total number of downloads, more understanding of the cause of outliers is desirable. A combined qualitative-quantitative approach seems the most promising, in which interviews with librarians at institutions is complemented with a more detailed analysis of the underlying usage patterns. A typical example was presented in Figure 2 in Section 4 identifying user sessions in which disproportionally large numbers of documents were downloaded shortly one after another from the same journal.

It must be noted that from the point of view of monitoring the use that users make of a publication archive, it is fully appropriate to include such downloads in statistics related to the archive and/or to the use individual users make of it. But from the point of view of research assessment focusing on performance or impact of documents, their authors and their institutions, it is questionable whether bulk downloads should be included in the counts. This illustrates that the type of data and of metrics to be included in a usage study very much depends upon the objectives of such a study. On the other hand, it must also be underlined that more insight is needed into the statistical effect that including or excluding of certain types of usage behavior has upon the outcomes of a quantitative analysis.

*Obsolescence functions and time delays per journal and type of document*

Large differences were found in download obsolescence rates among journals, subject fields, and types of document. It must be underlined again that the journals studied are *not* a random sample from the total population of journals in ScienceDirect. The aim of the selection was to include journals from different disciplines and cover all major disciplines and include sectionalized journals as well. The outcomes thus show how large the variability across journals and subject fields can be. Full length articles, reviews, short communications and editorial have different download obsolescence patterns; their differences are similar to those found for citations. The ratio of the number of the number of downloads per review to that per article is similar to the same ratio for citations. And short communications mature more quickly than full length articles both in terms of downloads and citations.

Applying a diachronous approach, the analyses at the journal level presented in the current paper show that, during the first 4 years after online publication date, *all* journals show a pattern in which the monthly number of downloads per document increases after their online publication date, reaches its peak after 2 to 8 months, depending upon the journal, and declines afterwards. This observation reveals large differences between journals in the age at which the download rate reaches its maximum. Equally important, it shows that the peak is reached one or more months *after* the online publication month. Such a behavior is qualitatively similar to that of citation obsolescence. This observation suggests that *both* processes are subjected to a delay. Following Parker (1982), Moed (2005) described the evolution of monthly download counts of a journal's documents as the sum of two exponential functions. Although the model showed a reasonable fit when applied to *Tetrahedron Letters*, a journal publishing short communications on a monthly basis with a relatively short life cycle, download obsolescence patterns per journal presented in Section 4 of the current paper revealed that a two factor model tends to be inappropriate as a standard.

Perhaps the most important conclusion that could be drawn is that the functional relationship of downloads obsolescence data strongly depends upon the way in which data are aggregated. When article and download counts are aggregated by year, differences in publication date among articles published within a year are ignored and do not play a role in the modeling of obsolescence functions. For instance, in the year of their publication, articles are followed on average only during a time period of 6 months rather than an entire year. In this respect, counts per month are more accurate, and reveal patterns that remain invisible when data are aggregated on a yearly basis. However, it must be noted that even in the monthly data presented in this paper the aggregation process of data can be assumed to have an effect, as it does not take count differences in publication dates of articles published in a month. The size of this effect can be assessed only if data on online publication date and download (or citation) date would be available on a daily basis. The authors of this paper plan to conduct a study based on daily counts in the near future.

*Skewness of downloads and citation article distributions*

Download counts tend to be less skewedly distributed among articles in a journal than citations, and, in agreement with this, the percentage of non-downloaded documents tends to much lower than the percentage uncited articles. More specifically, in the set of 62 study journals it was found for articles from 2008-2009 followed during 45 months that during this time period *all* documents are downloaded at least one, whereas 10 per cent of documents is not cited at all. Focusing on one single month, the 45[th] month after publication date, only 9 percent of documents is not downloaded anymore in that month. By contrast, 87 per cent is not cited in that month. In all 62 journals the 50 per cent most heavily cited articles account for a larger share of total citations to a journal than the share of downloads accounted for by the 50 percent most frequently downloaded documents. Differences between the share of downloads and citations thus accounted for range between 7 and 29 per cent, the largest differences tend to be found in mathematical and humanities journals, and the smallest ones in medical periodicals. From a purely statistical-analytical point of view, disregarding interpretational issues, download counts have a somewhat stronger position than citations: download counts tend to be two orders of magnitude higher and less skewedly distributed than citations.

*Statistical correlations between downloads and citations at the journal and article level*

Large differences in the degree of rank correlation between downloads and citations were found among subject fields at the journal level. The Spearman rank correlation coefficients varied between 0.3 for journals in *humanities* to 0.8 in chemical engineering and in *biochemistry and molecular biology*. Intuitively one might conjecture that subject fields in which the correlation is high tend to be very specialized fields, in which the readers of publications tend to be active researchers, or, in other words, fields in which the author and the reader populations tend to coincide. Fields in which the reader population is probably much wider than the research community – including for instance interested readers of articles in *humanities* and *social science* but who are active in other domains, or practitioners (engineers or nurses) using technical information from *engineering* and *nursing* journals – show a lower correlation. The analysis presented in this article do provide an indication of the validity of the above conjecture. But it must be noted that it did *not* measure the degree of overlap between author and user population, so that rigorous testing of the hypothesis that the degree of correlation between downloads and citation counts is positively related to this overlap, has to be been carried out in a follow-up study, due to a lack of information about the user or reader population.

The analysis also revealed large differences in the degree of correlation between downloads and citations among journals analyzed at the article level. The Spearman rank correlation coefficients vary between 0.30 for *Medieval History* and 0.80 for *Biochimica et Biophysica Acta – Gene Regulatory Mechanisms*. The outcomes are consistent with those obtained from the correlation analysis by discipline at the journal level. Comparing the Spearman rank correlation coefficient between downloads and citations *at the article level* for a journal from the 62 set with Spearman's Rho between the average download and citation rate *at the journal level* in the discipline(s) covered by that journal, it is found that these two correlation coefficients correlate positively themselves (both Pearson's R and Spearman's Rho amounts to 0.5). This outcome suggests that the differences between journals in the degree of correlation between downloads and citations at the article level are to some extent discipline-specific. Also, one can learn about average download and citation patterns of journals in a subject field by analysing download and citation counts at the article level in a representative journal covering that subject field. Following this hypothesis, the relatively deviant position in Figure 15 of four mathematics and one social science journals, which show a correlation at the article that is low compared to that at the journal level in their respective subject fields, should be attributed to the fact that these journals are not sufficiently representative for their subject fields.

Analyzing the NASA Astrophysics Data System, Kurtz et al. (2005b) found that "citations are a good predictor of downloads", but that "downloads are a poor predictor of citations". The results presented in the current paper do not confirm such a conclusion. There is perhaps even more evidence for the reverse conclusion, namely that downloads are a good predictor of citations and citations a poor or in any case a less valid predictor of downloads. Further research should throw more light upon this issue. It must be noted that Kurtz et al. studied mainly journals covering *one* single discipline, astronomy and astrophysics, whereas the journal set studied in the current article contains periodicals covering *all* domains of science and scholarship. Moreover, it applies a *diachronous* analysis following all articles from their online publication date during the same number of months, whereas Kurtz et al. adopt a *synchronous* viewpoint, studying downloads and citations made in one single year to articles published during a range of earlier years.

It was also shown that, even in journals in which downloads and citations strongly correlate, the articles appearing in the top of the citation ranking are not necessarily the most frequently downloaded ones, and vice versa. Comparing in the set of 62 study journals the top 10 most frequently cited documents with the set of the 10 most heavily downloaded ones, it was found that for 21 per cent of journals these two top sets have at most 2 documents in common, for 75 per cent at most 5, while for only 14 percent the overlap is 7 or more, and for none of the journals it is 9 or 10.

As outlined in Section 2, usage and citation leaks, differences between reader and author populations in a subject field, the type of document or its content, differences in obsolescence patterns between downloads and citations, and, last but not least, different functions of reading and citing in the research process, all provide possible explanations of a lack of correlation between download and citation counts and between rankings based on these counts.

**ANNEX A1: List of 62 journals analyzed in this article**

| Full Title |
| --- |
| Annals of Pure and Applied Logic |
| Applied Clay Science |
| Applied Ergonomics |
| Applied Surface Science |
| Behavior Therapy |
| Biochimica et Biophysica Acta - Bioenergetics |
| Biochimica et Biophysica Acta - Biomembranes |
| Biochimica et Biophysica Acta - Gene Regulatory Mechanisms |
| Biochimica et Biophysica Acta - General Subjects |
| Biochimica et Biophysica Acta - Molecular and Cell Biology of Lipids |
| Biochimica et Biophysica Acta - Molecular Basis of Disease |
| Biochimica et Biophysica Acta - Molecular Cell Research |
| Biochimica et Biophysica Acta - Proteins and Proteomics |
| Biochimica et Biophysica Acta - Reviews on Cancer |
| Cancer Letters |
| Child Abuse and Neglect |
| Design Studies |
| Differential Geometry and its Application |
| Earth and Planetary Sciences Letters |
| European Journal of Cancer |
| Fuzzy Sets and Systems |
| Journal of Applied Geophysics |
| Journal of Cultural Heritage |
| Journal of Dentistry |
| Journal of Econometrics |
| Journal of Economics and Business |
| Journal of Historical Geographpy |
| Journal of Hydrology |
| Journal of Informetrics |
| Journal of International Economics |
| Journal of Logic and Algebraic Programming |
| Journal of Medieval History |
| Journal of Phonetics |
| Journal of Science and Medicine in Sport |
| Journal of Wind Engineering and Industrial Aerodynamics |
| Limnologica |
| Lingua |
| Materials Science & Engineering A: Structural Materials: Properties, Microstructure and Processing |
| Materials Science & Engineering B: Solid-State Materials for Advanced Technology |
| Materials Science and Engineering C |
| Molecular Oncology |
| Ophthalmology |
| Performance Evaluation |
| Physica A: Statistical Mechanics and its Applications |
| Physica B: Condensed Matter |
| Physica C: Superconductivity and its Applications |
| Physica D: Nonlinear Phenomena |
| Physica E: Low-Dimensional Systems and Nanostructures |
| Phytochemistry |
| Phytochemistry Letters |
| Plant Physiology and Biochemistry |
| Plant Science Letters |
| Poetics |
| Powder Technology |
| Stem Cell Research |
| Surface Science |
| Tectonophysics |
| Tetrahedron Letters |
| Thin Solid Films |
| Topology and its Applications |
| Trends in Plant Science |
| Water Research |